\documentclass[preprint,aps]{revtex4}
\usepackage{graphicx}%
\usepackage{dcolumn}
\usepackage{amsmath}
\usepackage{longtable}
\begin{document}
\title
{Atomic and magnetic structures of (CuCl)LaNb$_2$O$_7$ and 
        (CuBr)LaNb$_2$O$_7$: Density functional calculations  }
\author
{ Chung-Yuan Ren$^{a,\dagger}$ and Ching Cheng$^{b}$}
\affiliation
{ $^{a}$ Department of Physics, National Kaohsiung Normal University, 
Kaohsiung 824, Taiwan  	\\
$^{b}$ Department of Physics, National Cheng Kung University, Tainan 701, Taiwan \\
$\dagger$ {\it E-mail address:} cyren@nknu.edu.tw  }
\begin{abstract}
The atomic and magnetic structures of (Cu$X$)LaNb$_2$O$_7$  
($X$=Cl and Br) are investigated using the density-functional calculations. 
Among several dozens of examined structures, an orthorhombic distorted 
$2\times 2$ structure, in  
which the displacement pattern of $X$ halogens resembles the model 
conjectured previously based on the empirical information is
identified as the most stable one. 
The displacements of $X$ halogens, together with those of Cu ions, 
result in the formation of 
$X$-Cu-$X$-Cu-$X$ zigzag chains in the two materials. Detailed analyses of the
atomic structures predict that (Cu$X$)LaNb$_2$O$_7$ crystallizes in the 
space group $Pbam$. The nearest-neighbor interactions 
within the zigzag chains are determined to be antiferromagnetic (AFM) for 
(CuCl)LaNb$_2$O$_7$ but ferromagnetic (FM) for (CuBr)LaNb$_2$O$_7$. 
On the other hand, the first two neighboring interactions between the Cu cations 
from adjacent 
chains are found to be AFM and FM respectively for both compounds.
The magnitudes of all these in-plane exchange couplings
in (CuBr)LaNb$_2$O$_7$ are evaluated to be about three times those in 
(CuCl)LaNb$_2$O$_7$. In addition,
a sizable AFM inter-plane interaction is found between the Cu ions 
separated by two NbO$_6$ octahedra. The fourth-neighbor 
interactions are also discussed.
The present study strongly suggests the necessity to go beyond the square $J_1-J_2$ 
model in order to correctly account for the magnetic property of (Cu$X)$LaNb$_2$O$_7$.	
\end{abstract}
\maketitle
\section{INTRODUCTION}
Low-dimensional quantum spin systems with 
frustrated interactions have drawn considerable attention for several 
decades \cite{book1}. In particular, the 
square-lattice $S=1/2$ frustrated Heisenberg magnets with 
first-neighbor exchange constant $J_1$ and  
second-neighbor constant $J_2$ are increasingly interesting 
due to their unusual ground states and quantum phenomena
\cite{DM}-\cite{SMS}. Based on the $J_1-J_2$ 
model studies, there exist several phases as a function of
$J_2/J_1$. When $J_1$ dominates or $J_2$ is ferromagnetic (FM),
the system is either N\'{e}el antiferromagnetic (NAFM) or FM
depending on the sign of $J_1$ (Refs. \cite{CCL,SOW,SSPT,SMS}). 
When $J_2$ is antiferromagnetic (AFM) and dominates, 
there appears the so-called columnar AFM (CAFM) order \cite{CCL,SOW}
with antiferromagnetically coupled FM chains.
%characterized by magnetic wave vectors $Q=(\pi,0)$ or $(0,\pi)$. 
The CAFM and FM, or CAFM and NAFM ordered phases are separated by
the intermediate quantum-disordered phases, 
the nature of which is not yet fully resolved 
\cite{DM,BFP,SWHO,CBPS,SOW,SSPT,SMS}.
The recent discoveries of quasi-two-dimensional materials are 
realizations to test the validity of the $J_1-J_2$ model. 
Prominent among them are Li$_2$VO(Si,Ge)O$_4$ (Refs. \cite{MS,MCLMTMM}),
$AB$(VO)(PO$_4$)$_2$ ($A,B$=Pb, Zr, Sr, Ba) 
(Refs. \cite{KPhD,KKG,KRSSG}),
(CuBr)$A^{'}_2B^{'}_3$O$_{10}$ ($A^{'}$=Ca, Sr, Ba, Pb; $B^{'}$=Nb, Ta)
(Ref. \cite{TKBKYNK}),
and (Cu$X$)LaNb$_2$O$_7$ ($X$=Cl, Br) (Refs. \cite{YOTYIKKAY,YOTKKAY}). 
(Cu$X$)LaNb$_2$O$_7$ compounds are of particular interest because they 
allow systematic tuning and understanding of 
the structural and magnetic properties, which are  
plausibly connected with the phenomenon of 
high-$T_c$ superconducting cuprates. 

Although divalent copper with the electronic configuration $d^{9}$ 
should be Jahn-Teller active and lead to the cooperative 
lattice distortion (e.g., perovskite KCuF$_{3}$ (Ref. \cite{LAZ})), 
the precise crystal structure of the layered
copper oxyhalides (Cu$X$)LaNb$_2$O$_7$ is still under debate.
Earlier structural studies on (Cu$X$)LaNb$_2$O$_7$ were carried out 
with the tetragonal space group $P_4/mmm$, where the 
Cu and $X$ sites possess the $C_4$ symmetry \cite{KKZW,KLZCSFSOW}.
While the Rietveld refinement gave satisfactory results, 
the thermal parameter for halogens remained large. Besides, in this 
structure copper is 
in a significantly squeezed octahedral coordination with two 
short Cu-O bonds (about 1.9 \AA) and four rather long Cu-$X$ bonds (2.7 \AA), 
which  are also quite unusual.
Subsequently, the neutron diffraction experiment \cite{CKW} 
proposed that the Cl ions in (CuCl)LaNb$_2$O$_7$ ((CuCl)LNO) shifted away  
from the ideal Wyckoff $1b$ position \cite{KLZCSFSOW}. The transmission
electron microscopy measurement on (CuCl)LNO (Ref. \cite{YOTYIKKAY})  
revealed superlattice reflections corresponding to 
an enlarged $2\times 2$ unit cell. The nuclear magnetic
resonance and the nuclear quadrupole resonance experiments for (CuCl)LNO and  
(CuBr)LaNb$_2$O$_7$ ((CuBr)LNO)
further demonstrated the lack of the tetragonal symmetry at 
both Cu and Cl/Br sites \cite{YOTKKAY,YOTYIKKAY}.

The magnetic properties of (CuCl)LNO and (CuBr)LNO are also unusual and 
lack a clear microscopic interpretation. The former exhibits a 
spin liquid phase with a spin gap \cite{KKONNHVWYB,YOTYIKKAY,KYKTNHKBOAY} 
that are incompatible \cite{WD} with the square 
$J_1-J_2$ model. On the other hand,
it has been reported \cite{OKKYBNHNHKSAY} that the replacement of Cl by Br 
leads to a CAFM order in (CuBr)LNO at low temperatures. 
However, it is unclear whether the Cu ions 
connected with the dominant exchange interaction couple 
ferromagnetically or antiferromagnetically \cite{YOTKKAY}.
Moreover, both (CuCl)LNO and (CuBr)LNO are claimed to be FM $J_1$  
compounds \cite{KKONNHVWYB,OKKYBNHNHKSAY} whose
justifications largely rely on the $J_1-J_2$ model. Yet, the structural 
study \cite{YOTKKAY} raised doubts over the validity of the model.
Therefore, unambiguous determination of the 
crystal structure is crucial for understanding these
complex systems. 

At present, there are several structural models proposed for the 
Cu$X$ plane. Whangbo and Dai \cite{WD} suggested a model that 
consists of different ring clusters to explore the exchange couplings. 
However, the existence of inequivalent Cu and Cl sites in such a model
is in contradiction to the experimental results that both Cu and $X$ 
occupy a unique crystallographic site with no substantial disorder 
\cite{YOTYIKKAY,YOTKKAY}. Yoshida {\it et al.} \cite{YOTYIKKAY}, 
based on the empirical evidence, proposed an orthorhombic distorted 
$2\times 2$ structure (hereafter referred to as the YY model). 
In this model, the displacement of Cl ions 
generates different exchange couplings among the nearest neighboring Cu pairs. 
A Cu dimer formed 
by the dominant exchange interaction was considered \cite{YOTYIKKAY} 
to study the spin-gap behavior. The same structural model 
was shown \cite{YOTKKAY} to consistently account for (CuBr)LNO. 
The third model, suggested by Tsirlin and Rosner (TR) \cite{TR} is also
characterized by an ordering pattern but with a $2\times 1$ periodicity,   
where the local environment of copper  
is distorted to form the CuO$_2$Cl$_2$ plaquette.   

First-principles calculations have proven to be an appealing method   
to deal with complex systems \cite{Er,MV,RSZODP}.
Such a method can efficiently and reliably calculate the total energy, 
which is crucial in determining the most stable structure in order to study 
all relevant physical properties. In this work, we will investigate the 
atomic structure and resultant magnetic property of (CuCl)LNO and (CuBr)LNO 
based on the density functional theory. Our results show that, among 
several dozens of examined structures, the distortion pattern of the 
most stable one is similar to that of the YY model.  
The displacement of the $X$ ions changes the environment 
of copper to form the CuO$_2$$X_2$ plaquette. In addition, these two
materials crystallize in the space group $Pbam$. 
The FM chains in CAFM (CuBr)LNO are found to be
along the direction which is contrary to the previous conjecture
\cite{YOTKKAY}. It will be shown that (CuCl)LNO still belongs to 
the AFM $J_1$ compound. 
The first- and second-neighbor exchange couplings
of (CuCl)LNO and (CuBr)LNO are also discussed in detail.
\section{CRYSTAL STRUCTURE AND COMPUTATIONAL DETAILS}
Figure \ref{fig1} illustrates the basic 
crystal structure of the copper oxyhalides
(Cu$X$)LaNb$_2$O$_7$. It is made up of copper-halogen planes  
and nonmagnetic double-perovskite LaNb$_2$O$_7$ slabs. The La ions are located
at the 12-coordinate sites of the double-perovskite slabs. The Cu$X$ planes 
and the LaNb$_2$O$_7$ slabs alternate along the $c$ direction such that 
the copper is six-fold coordinated, bridging between the apical 
O ions of NbO$_6$ octahedra and surrounded by four $X$ halogens. 
Because of the short Cu-O bond length ($\sim$1.9 \AA), 
the Cu$X$ plane is more appropriately considered as a Cu$X$O$_2$ layer. 
The initial structural study on (Cu$X$)LaNb$_2$O$_7$
was carried out with the space group $P_4/mmm$, where both Cu and $X$ 
have the $C_4$ symmetry \cite{KKZW,KLZCSFSOW}. 
In this model (hereafter referred to as C4), 
the Cu and $X$ ions are located
at the Wyckoff $1d$ and $1b$ positions, respectively (Fig. \ref{fig2}(a)).
Later studies \cite{CKW,YOTYIKKAY} proposed that Cl ions are 
displaced from the $C_4$-symmetry positions. 
The YY $2 \times 2$  model is represented in Fig. \ref{fig2}(b). 
The displacement of $X$ ions on the Cu$X$ plane leads to the formation of
the $X$-Cu-$X$-Cu-$X$ zigzag chains, as indicated in Fig. \ref{fig2}(b). 
The original equivalent and perpendicular Cu chains are now distinguishable. 
Here, the direction extending along the zigzag chains 
is defined as the $b$ axis. 
 
The present calculations were based on
the generalized gradient approximation (GGA) \cite{PBE} 
to the exchange-correlation energy functional of the density functional theory.
It is known \cite{ZR,CS,MG}
that Cu-derived oxide compounds are usually strongly correlated systems.
The correlation effect is important for the present systems to
understand their ground state. Therefore, 
the on-site Coulomb interaction $U$ for Cu $3d$ electrons was also included
\cite{LAZ} (GGA+$U$) in this work. 
Since the on-site exchange interaction $J$ is expected 
to be less influenced by the solid state effects \cite{TR}, the relation
 $J=0.1U$ was used \cite{SHT} for different choices of $U$. 
The projector-augmented-wave potentials, as implemented in
VASP \cite{KJ,KF}, were employed for the interactions between the 
ions and valence electrons.
The plane-wave basis set with an energy cut-off of 500 eV was used. 
To minimize numerical uncertainties, structural optimizations were performed 
using a $2\times2$ supercell for all the test structures 
unless specified otherwise.
The $6\times6\times4$ Monkhorst-Pack grids were taken
to sample the corresponding Brillouin zone. 
The lattice parameters and  atomic positions
were relaxed until the total energy
changed by less than $10^{-6}$ eV per conventional cell
and the residual force was smaller than 0.01 eV/\AA.
\section{RESULTS AND DISCUSSION}
\subsection{Energetics}
We first calculated the total energies of the C4 structure and several 
$2\times1$ and $2\times2$ distorted structures with different displacement 
patterns of 
the $X$ halogens. The YY $2\times2$ model is found to be the most stable one. 
As compared to the C4 structure, the YY model has 
a significant 0.3 and 0.2 eV/fu lowering in the energy of 
(CuCl)LNO and (CuBr)LNO, respectively. This directly rules out the possibility that 
the two compounds crystallize in the $C_4$ symmetry. 
Particularly, over the full $U$ range from 0 eV to 8 eV, 
the YY model is 0.1 eV/fu 
lower than the TR $2\times1$ model (see Fig. 4 in Ref. \cite{TR}) 
for both materials. The nonmagnetic calculations with and without $U$  
lead to the same conclusion. Therefore, {\it the structural distortion
outweights the magnetism and on-site correlation effects}
in determining the atomic structure. Note that,
besides Refs. \cite{YOTYIKKAY} and \cite{YOTKKAY}, very recent experimental 
evidences \cite{KTKANNKIUUY} also confirm that the original unit cell 
should be double along both the $a$ and $b$ axes for the family of these 
compounds. Our study therefore provided theoretical support for the 
stabilization of the YY $2\times2$ model.

To examine whether there exists other more stable
structure with the $X$ ions restricted to the
Cu plane, we perform the calculations for (CuCl)LNO
with twenty sets of random displacements of all four Cl ions from the
positions in the YY model. However, no such structure 
was found. The resultant configurations of the trial structures 
are either relaxed back to the 
YY model or trapped into a nearby higher energy minimum.

Next, we allow the halogens in the YY model to move off the 
plane. It is found that the $X$ ions in the relaxed structures 
are $0.02-0.04$ \AA~ away from the Cu plane. However, the change in the
total energy is rather small. At $U$=0 and 8 eV, the results of 
both materials show 
that the energy differences are only within 2 meV per
$2\times2$ supercell while the energies for the structures with 
the $X$ ions fixed in the 
plane remain lower. We also examine the two structures in 
Figs. 16(e) and 16(f) of Ref. \cite{YOTYIKKAY}, 
which are based on another $2\times 2$ configuration 
with the Cl ions displaced away from the Cu plane in a different way. 
The calculations indicate that, after relaxation,
both are energetically about 0.2 eV/fu higher than the YY model. 
The increase in the total energy mostly comes from the different in-plane 
structural distortions. Again,
the contribution from the $z$-component shift of 
Cl ions is rather minor. 
Hence, the distortion on the Cu$X$ plane is predominantly crucial to 
stabilize the atomic structure. In the following discussion, 
we shall focus on the YY model with $X$ ions kept in the Cu plane \cite{com1}. 

Now, we analyze the total energies influenced by the on-site 
Coulomb interaction and the different magnetic configurations 
shown in Fig. \ref{fig3}. Here, SC1 and SC3 are FM and NAFM. SC2 and SC4
are both CAFM, with the FM chain along the $b$ and $a$ directions, 
respectively. The results are displayed in Fig. \ref{fig4}, where  
the energy of SC2 was chosen as a reference. 
For (CuCl)LNO, the energies of SC1, SC2, and SC4
are very competing. The differences among them are within 1 meV/fu 
when $U\geq 6$. The first two
are even almost identical around $U$=4 eV. Clearly, Fig. \ref{fig4}(a) shows 
that the SC3 is the lowest energy state and its energy 
is well separated from those of the other three magnetic structures. 
In the (CuBr)LNO case, similar 
tiny energy differences but between SC1, SC2, and SC3 are also found. 
Interestingly, when
the FM chain in the CAFM state is set parallel along the $a$ axis, as in SC4, 
the total energy over the examined $U$ range is much higher than those of 
the other three configurations, indicating that the Cu ions along the $a$-axis 
should not couple ferromagnetically.
This finding is contrary to the previous conjecture \cite{YOTKKAY}. 
The different energy ordering for the four magnetic configurations
between the two compounds are
conceivable since the magnetic interactions through the path Cu-$X$-Cu 
depend subtly on the small structural variation via the $X$-ion size 
effect. We will return to this issue in Sec. IIID when considering the  
various exchange couplings.
\subsection{Atomic structure}
Tables \ref{tab1} lists the fully optimized structural parameters 
of both materials. 
For comparison, those obtained by the C4 model are also included.
As can be seen in this table, the evaluated lattice constants are
in good agreement with the experimental data \cite{CKW,KKZW}.
The discrepancies between them are only within 
$ 1\%$, the typical errors in the density-functional calculation. 
The $a$ and $b$ lattice constants of (CuCl)LNO are smaller than those of 
(CuBr)LNO, which is due primarily to the size effect of Br in the layered
structure. 

To discuss the structural distortion, we take the (CuCl)LNO case as an example. 
In the C4 model, copper is in the  
squeezed octahedral coordination with four long Cu-Cl 
bonds [$d$(Cu-Cl)=2.77 \AA] and two short Cu-O  
bonds [$d$(Cu-O)=1.85 \AA]. The displacements of the Cl and Cu ions 
in the YY model lead to two shorter Cu-Cl bonds of 2.38 and 2.39~\AA,
forming the Cl-Cu-Cl-Cu-Cl zigzag chain to stabilize the structure. 
The rest two Cu-Cl interatomic distances are increased to $3.27-3.29$~\AA.
In particular, the Cu-O bond length remains short 
after the structural distortion (from 1.85 \AA~to 1.88 \AA), 
indicating the strong 
bonding character between Cu and O ions.
The calculated interatomic distances are comparable to those reported 
previously \cite{CKW}.
As a result, the distortion yields the nearly planar 
CuO$_2$Cl$_2$ rather than the octahedral CuO$_2$Cl$_4$ environment around 
the Cu ion (Fig. \ref{fig5}(a)). 
The resultant CuO$_2$Cl$_2$ planar structure is reminiscent of 
the conventional 
CuO$_4$ plaquette, which is commonly observed in copper oxides, 
e.g., La$_2$CuO$_4$ (Ref. \cite{CS})  and Sr$_2$CuO$_3$ (Ref. \cite{MG}). 
It should be noted that the CuO$_2$Cl$_2$-plaquette zigzag chains
was also reported in the TR model \cite{TR}. 
Additionally, the basic electronic structure is similar to that of 
the CuO$_4$ planar unit, which will be demonstrated in the next section. 
Combined with the energetic advantage mentioned above, 
the YY model provides a realistic description for the atomic
structures of (CuCl)LNO and (CuBr)LNO .   

From a closer analysis of the positions of all ions in (CuCl)LNO, we found that
the distorted atomic structure in the YY model belongs to the space group
{\it Pbam} (No. 55) \cite{Book0}. 
The atomic positions are summarized in Table \ref{tab2}.
Clearly, the deviations of the Cl ions 
from the $C_4$-symmetry positions are as large as 0.66 \AA, and these values 
are four times larger than 
those of Cu ions. Note that the displacements along the $a$ axis are
more significant than those along the $b$ axis for both ions to form the 
zigzag chains. 

As expected, the structural distortion on the CuCl plane leads La, Nb 
and O ions to shift from the the $C_4$-symmetry positions. Figure 
\ref{fig5}(b) and Table \ref{tab2} show the significant tilting and 
distortion of NbO$_6$ octahedra. Such a tilting distortion is typical for
perovskite oxides structures \cite{Book1}. 
Particularly, La ions shift along the $b$ axis by an amount of 0.10 \AA.  
This displacement of La from the $C_4$-symmetry positions well agrees with 
the experimental nonzero value of the EFG tensor at La sites \cite{YOTYIKKAY}, 
a strong evidence for the structural distortion in (CuCl)LNO. 
We also found that Nb
ions shift along the $a$ axis by a relatively smaller amount of 0.02 \AA. 
Note that the $a$ ($b$) component of La (Nb) displacement is almost negligible.

Taking into account the tilting of  
the NbO$_6$ octahedra in the (CuCl)LNO is important 
for providing a realistic description of the distortion on the CuCl plane. 
Figure \ref{fig5}(c), the top view of the atomic structure, clearly 
demonstrates that  
the cooperative tilting of the NbO$_6$ octahedra in the space group $Pbam$ 
results in a $2\times 2$ periodicity and leads to 
the zigzag chains with the same 
periodicity. It is worth pointing out that the higher symmetric 
$2\times 1$ zigzag chains in the TR model were investigated without 
consideration of the effect due to the tilting of the NbO$_6$ octahedra, where
these octahedra were still kept at the C4 tetragonal sites \cite{TR}. 
Allowing the
tilting distortion of NbO$_6$ octahedra in (CuCl)LNO lowers the symmetry of the 
atomic structure and correspondingly that of the zigzag chains and therefore 
leads to a lower total energy. In the YY model, the zigzag chains have 
the glide symmetry about $u=1/4$ and $v=1/4$. Specifically,
the Cu-Cl bond
of 2.39 \AA~ in the zigzag chain is next to the Cl-Cu bond of 2.38 \AA~ in the
adjacent chain and vice versa. As compared to those in the TR model, such a 
{\it complementary} arrangement between adjacent chains in the YY model allows
a further lowering in energy.  Now, it is evident that the 
YY $2\times 2$ model in the present study is energetically more stable 
than the TR $2\times 1$ one.   
Similar conclusion holds for (CuBr)LNO.
\subsection{electronic structure}
Figure \ref{fig6} depicts the orbital- and site-projected
densities of states (DOS) of (CuCl)LNO with $U$=6 eV, where the
valence-band maximum ($E_v$) is set to zero. The orbitals are 
projected in the local coordinates with the $x$ and $y$ axes directed to
the neighboring Cl ions and the  $z$ axis coinciding with the crystal 
$c$ axis (Fig. \ref{fig2}(b)). Among the major valence state
region of 6.3 eV, the higher-energy part consists almost exclusively of O and Cl
$p$ states. There is larger contribution from the Cl $p$ state just below $E_v$. 
The lower-energy part, dominated by the Cu $d$ states, is splitted into the doubly
occupied $dxy$, $dyz$, $dzx$, $d(x^2-y^2)$, and singly occupied 
$d(3z^2-r^2)$ states. As compared to the GGA DOS (not shown
here), the GGA+$U$ shows an essential redistribution of the Cu $3d$ DOS, 
i.e., from being above to below the O and Cl $p$ states. 
That the energy gap ($E_g$) lies between the occupied anion $p$ states and 
the unoccupied Cu $d$ states is similar to those in the charge-transfer 
insulators, e.g.,
La$_2$CuO$_4$ (Ref. \cite{CS}) and Sr$_2$CuO$_3$ (Ref. \cite{MG}).
The sharp peak of the low-lying Cu $d(3z^2-r^2)$ state is a result of
the strong bonding between the Cu and O ions with a considerably short Cu-O 
bond length of 1.88 \AA~ (see Table \ref{tab1}). Note that the 
$d(3z^2-r^2)$ orbital was hybridized with little Cl $p$ component.
The on-site correlation $U$ leads to the 
half-filling of the $d(3z^2-r^2)$ orbital, of which the lobes 
point to the O ions. These results imply a single orbital ground 
state. Figure \ref{fig6} shows that, 
due to the hybridization with the O $p$ state \cite{CS}, the 
$d(3z^2-r^2)$ bonding-antibonding separation (8.3 eV) is larger than the 
value of $U$. 
Therefore, the electronic structure due to 
the CuO$_2$Cl$_2$ plaquette in the YY model is very similar to those of  
other copper oxides \cite{CS} with planar CuO$_4$ units.

We found that the structural distortion and magnetism together already 
open up the gap. The $E_g$ of (CuCl)LNO obtained by the GGA is 0.3 eV. However, 
this result is insufficient to account for the green color
appearance \cite{KKZW} of this compound. At $U=6$ eV, the $E_g$ is increased 
to 1.8 eV. Further increase of $U$ makes no  
significant widening for the band gap. 
The main structures in the DOS of (CuCl)LNO are also found in that of (CuBr)LNO, 
except for the smaller $E_g$ of 1.5 eV. At this choice of $U$, the
local magnetic moment at the Cu site of (CuBr)LNO is evaluated to be 
0.6 Bohr magneton, 
which agrees with the experiments \cite{YOTKKAY,OKKYBNHNHKSAY}.
An amount of 0.1 Bohr magneton at Br sites is also observed.
Hence, we choose the optimal $U$ = 6 eV case to
discuss the corresponding atomic and electronic properties.
\subsection{exchange interaction}
Finally, we discuss the exchange couplings for both (CuCl)LNO and (CuBr)LNO. 
In the undistorted C4 structure, the interactions between the Cu ions
can be approximately modeled by the Heisenberg Hamiltonian 
\(\hat{H}=J_1\sum_{NN}S_i\cdot S_j +J_2\sum_{2NN}S_i\cdot S_j\),
where the sums run over the first and second nearest-neighbor pairs, 
respectively, and $S_i$ corresponds to the spin moment at site $i$. 
The relevant exchange couplings can be then determined by applying the
model to the calculated energies of different spin configurations. 
For the YY model, the formation of the $X$-Cu-$X$-Cu-$X$ zigzag chains 
along the $b$ axis (Fig. \ref{fig2}(b)) lifts the
tetragonal symmetry and leads to inequivalent superexchange pathways, 
as indicated in Fig. \ref{fig2}. As a result,  
the $J_1$ in the C4 structure is split into  
$J_{11}$, $J_{12}$, and $J_{13}$, with the former two 
now being the first neighboring inter-chain interactions and
the latter the first neighboring intra-chain interaction.
The original $J_2$ coupling is split into two 
inequivalent $J_{21}$ and $J_{22}$, which are correspondingly the 
second neighboring inter-chain interactions. We investigate these 
interactions via the various spin configurations in Fig. \ref{fig3}. 
The results are summarized in Table \ref{tab3}. 

We first discuss the results from the C4 model. Table \ref{tab3} shows that
$J_1$ is almost negligible as compared to $J_2$. This is reasonable because,
as illustrated in Fig. \ref{fig7}(a), there is no overlap between the 
Cu2 $d(x^2-y^2)$ and $X$4 $p$ orbitals. Therefore, even with the obvious
overlapping of the Cu1 $d$ and $X$4 $p$ orbitals, Cu1 and Cu2 could hardly 
interact with each other. On the other hand,  
Cu1 can interact with Cu3 via the $X$4 $p$ orbital. 

Based on the $J_1-J_2$ model, both (CuCl)LNO and (CuBr)LNO were 
previously claimed \cite{KKONNHVWYB,OKKYBNHNHKSAY} 
to be FM $J_1$ magnets with competing AFM $J_2$ interactions, as
in the case of Pb$_2$VO(PO$_4$)$_2$ (Ref. \cite{KRSSG}). 
Table \ref{tab3} 
indeed shows that $J_1<0$ and $J_2>0$ for both materials in the C4 
model, a direct consequence of the Hund's coupling and virtual electron hopping. 
However, the recent structural study \cite{YOTKKAY}
has raised serious doubt over the validity of the $J_1-J_2$ model in such 
materials. Our calculations also indicate that consideration 
of the structural distortion leads to the opposite results. For (CuCl)LNO, 
the effective interactions 
$(J_{11}+J_{12}+2J_{13})/4$ and $(J_{21}+J_{22})/2$ in the YY model 
are found to be AFM and FM, respectively. And they both become 
FM for (CuBr)LNO at large $U$s. These results come from the complicated
interplay between the Hund's coupling,
virtual electron hopping, the distorted structure and $X$-ion size effect.
It should be noted that the TR model \cite{TR} also results in 
a leading AFM coupling in (CuCl)LNO. 
This implies that the simple $J_1-J_2$ model is unable to describe 
the present systems. Moreover, the first neighboring interactions become 
more significant as compared to the second neighboring ones. 
Figure \ref{fig7}(b) clearly shows that, unlike the C4 case, 
$\angle$ Cu1-$X4$-Cu2 is no longer 90$^o$ (Table \ref{tab1}) due to the 
structural distortion. This will lead to the overlapping of 
Cu2 (Cu4) $d$ and $X$4 $p$ orbitals, and enhance the interaction between 
Cu1 and Cu2 (Cu4). 

In fact, Fig. \ref{fig8}(a) shows that for (CuCl)LNO, 
$(J_{11}+J_{12})/2>0$, $J_{13}>0$, and $(J_{21}+J_{22})/2<0$  
for all the $U$s considered. 
It is now clear that, since the interactions due to 
all the corresponding spin pairs in SC3 satisfy these conditions, 
(CuCl)LNO in SC3 is much more stable than in the rest 
three configurations of Fig. \ref{fig3}. Actually, SC3  
is the most stable structure among all the 
spin configurations with the interactions up to second-nearest neighbors. 
However, by comparing 
Fig. \ref{fig8}(b) with Fig. \ref{fig8}(a), we found that the
$J_{13}$ in (CuBr)LNO becomes FM. This could be ascribed to the 
fact that the interaction $J_{13}$
depends sensitively on the angle of the Cu1-$X$4-Cu2 superexchange path in
Fig. \ref{fig7}, and the replacement of Cl by Br will change this angle 
and modify the 
interaction. Therefore, for (CuBr)LNO, the first neighboring couplings
within the chains are FM, and the first two neighboring couplings
between adjacent chains are AFM and FM, respectively.
None of the four structures
in Fig. \ref{fig3} satisfies these conditions. Specifically,  
the interactions due to the corresponding spin pairs in SC4 are 
all opposite to these couplings, giving rise to the result of 
the SC4 being the highest-energy structure for (CuBr)LNO. 

The above analysis seems to indicate that, contrary to 
previous expectations \cite{KKONNHVWYB,YOTYIKKAY,KYKTNHKBOAY,OKKYBNHNHKSAY}, 
(CuCl)LNO rather than (CuBr)LNO is less frustrated \cite{OKKYBNHNHKSAY}. 
To check 
the reliability of our calculations, we perform the total-energy calculation
for an additional structure SC5 with three of the four spins being the same but 
opposite to the fourth one. For all the possible choices of four magnetic
structures containing SC5 in solving the Heisenberg Hamiltonian, the deviations 
of the relevant couplings (dashed lines in Fig. \ref{fig8}) 
from those obtained by SC1$-$SC4 are less than 1.0 meV. More importantly,  
the signs of these interactions remain unaltered for both materials. 

One plausible explanation for the above puzzling discrepancy is that 
the third neighboring couplings between 
different zigzag chains \cite{com3} may not be completely 
negligible \cite{TR,KKONNHVWYB}. Actually, in the YY model,
the two CuO$_2$$X_2$ plaquettes with the Cu1 and Cu5 ions
in Fig. \ref{fig7}(b) could be considered approximately coplanar. 
Kageyama {\it et al.} \cite{KKLW} argue that this kind of coplanarity provides 
an opportunity for the interaction between Cu1 and Cu5  
through the overlap of the
Cu1 $d(x^2-y^2)-$X4 $p-X$5 $p-$Cu5 $d(x^2-y^2)$ orbitals.
So, from the structural geometry point of 
view, such a long-range coupling (8.5 \AA)
could be possible. However, to examine and 
identify these couplings, one has to 
take into account additional eight inequivalent couplings and
use a larger supercell whose 
corresponding calculations are very time-consuming and yet, likely, not accurate
enough for the present purposes (Fig. \ref{fig4}). Therefore, we will
not address this issue presently. Further theoretical
and experimental work is required to clarify this point.

It is worth pointing out that the first neighboring interactions 
within the chains and between the adjacent chains have the opposite 
signs for (CuBr)LNO. The spatial asymmetry of these results 
again demonstrates the inappropriateness of the square $J_1-J_2$ model for 
the Br compound. Furthermore, for both materials,
the couplings between the adjacent chains 
are very competitive to the intra-chain interactions, sharply 
contrary to the previous conclusion \cite{TR}.  Tsirlin and Rosner
have argued \cite{TR} that in the TR model, where the basic structure
element is also the CuO$_2$Cl$_2$-plaquette zigzag
chain, the large hopping runs 
along the chain and leads to the strongest interaction. 
According to their discussion,
the inter-chain interaction is rather weak due to the long 
"nonbonding" Cu-Cl distance and the lack of the proper superexchange 
path. In the present study, despite the similar backbone in the YY model, 
the couplings between the adjacent chains are shown to be still 
substantial. As mentioned before, 
the strength of exchange interactions between two 
spin sites should be determined by the overlap of orbitals rather than the 
distance between them. The interactions between the Cu ions from adjacent 
chains could be significant through the path mediated 
by the extended $3p$ orbital of Cl ions (vs O$^{2-}$)
and would be enhanced in the Br case with the further extended $4p$ orbital.  
Indeed, our calculations show that 
all the in-plane exchange couplings 
in (CuBr)LNO are three times larger than those in (CuCl)LNO.

For the inter-layer interaction $J_{\perp}$, 
Table \ref{tab3} shows that the $J_{\perp}$ is AFM, 
in agreement with the experiment \cite{OKKYBNHNHKSAY}. 
When compared to the in-plane interaction, the $J_{\perp}$ in (CuCl)LNO is 
non-negligible, implying that some long-path (12 \AA) interaction between 
the Cu ions is still cooperative.
The origin of this long-range coupling could be associated with 
the interaction through
the Cu $d(3z^2-r^2)$ orbital. As discussed in Sec. IIIC, this orbital strongly 
overlaps with the O $p_z$ orbital. The O $p_z$ orbitals further couple
with Nb $4d$ orbitals. Therefore, the inter-plane coupling $J_{\perp}$ 
shall involve the Cu-O-Nb-O-Nb-O-Cu path. In (CuBr)LNO, however, the coupling 
is found to be relatively less significant.
The strength of all the interactions interested here is decreased with increasing 
$U$. The evolution is expected since adding $U$ makes 
the wavefunctions more localized and the virtual electrons
hopping less favorable. 
\section{CONCLUSIONS} \label{con}
In conclusion, we have investigated the atomic structure and  
magnetic property of the copper oxyhalides 
(Cu$X$)LaNb$_2$O$_7$($X$=Cl and Br) based on the  
density functional theory. The calculations show that, among 
the examined structures, the YY $2\times 2$ 
model proposed by Yoshida {\it et al.}  
\cite{YOTYIKKAY} has the lowest energy. This model is 
significantly more stable than both the undistorted 
C4 structure and the TR $2\times 1$ model suggested recently by
Tsirlin and Rosner \cite{TR}. 
The $X$ and Cu ions in the YY model are displaced to form the 
$X$-Cu-$X$-Cu-$X$ zigzag chains and the local environment of copper  
is distorted to form a nearly CuO$_2$$X_2$ plaquette. 
We found that (Cu$X$)LaNb$_2$O$_7$ crystallizes in the space group $Pbam$. 
The cooperative tilting of the NbO$_6$ octahedra leads to the lower symmetry of
the zigzag chains with a $2\times 2$ periodicity.  With consideration of
the on-site Coulomb interaction, the YY model shows the single-orbital scenario 
typical for copper oxides and oxyhalides. 

\begin{figure}[h]
        \caption{ } \label{fig1}
\includegraphics{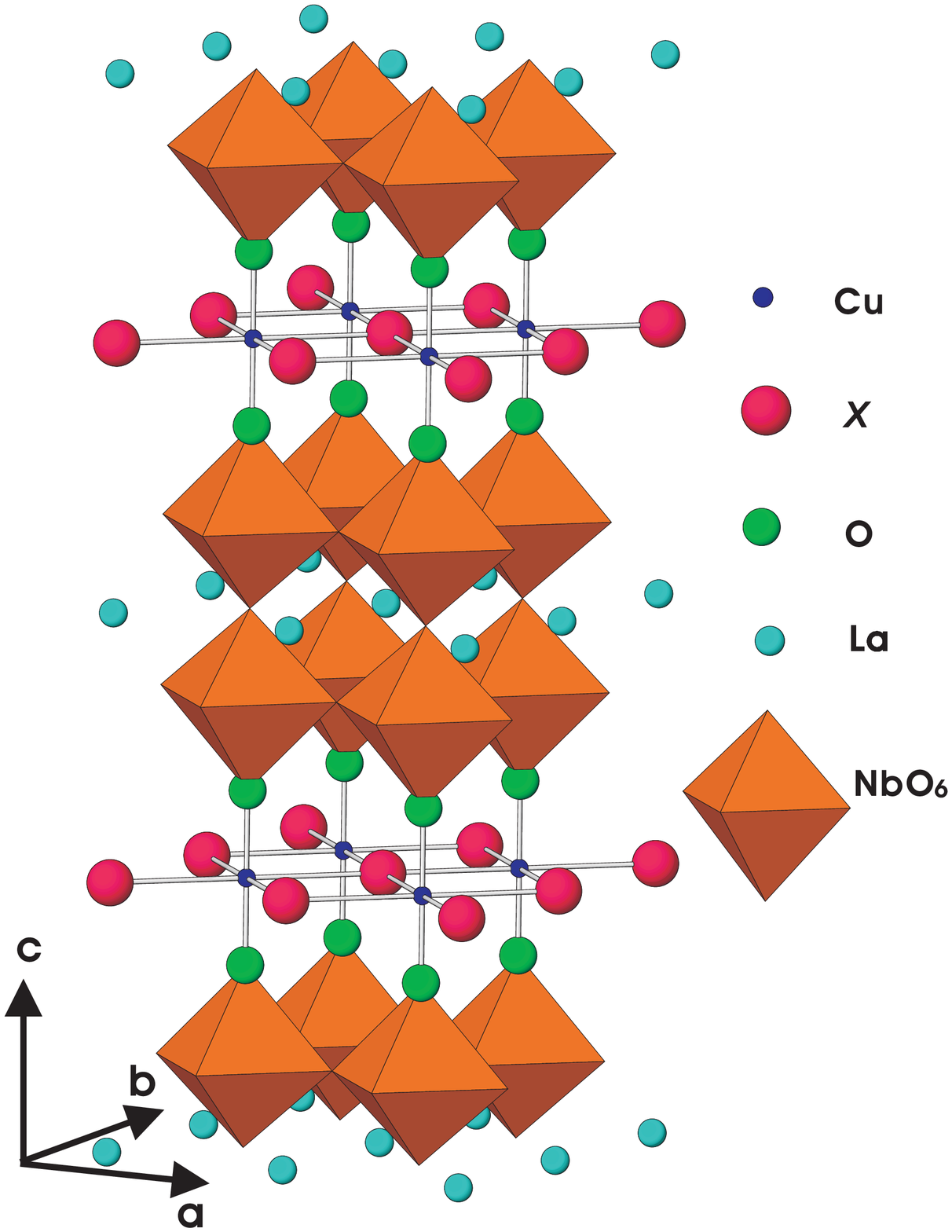}
\end{figure}

\begin{figure}[h]
        \caption{ } \label{fig2}
\includegraphics{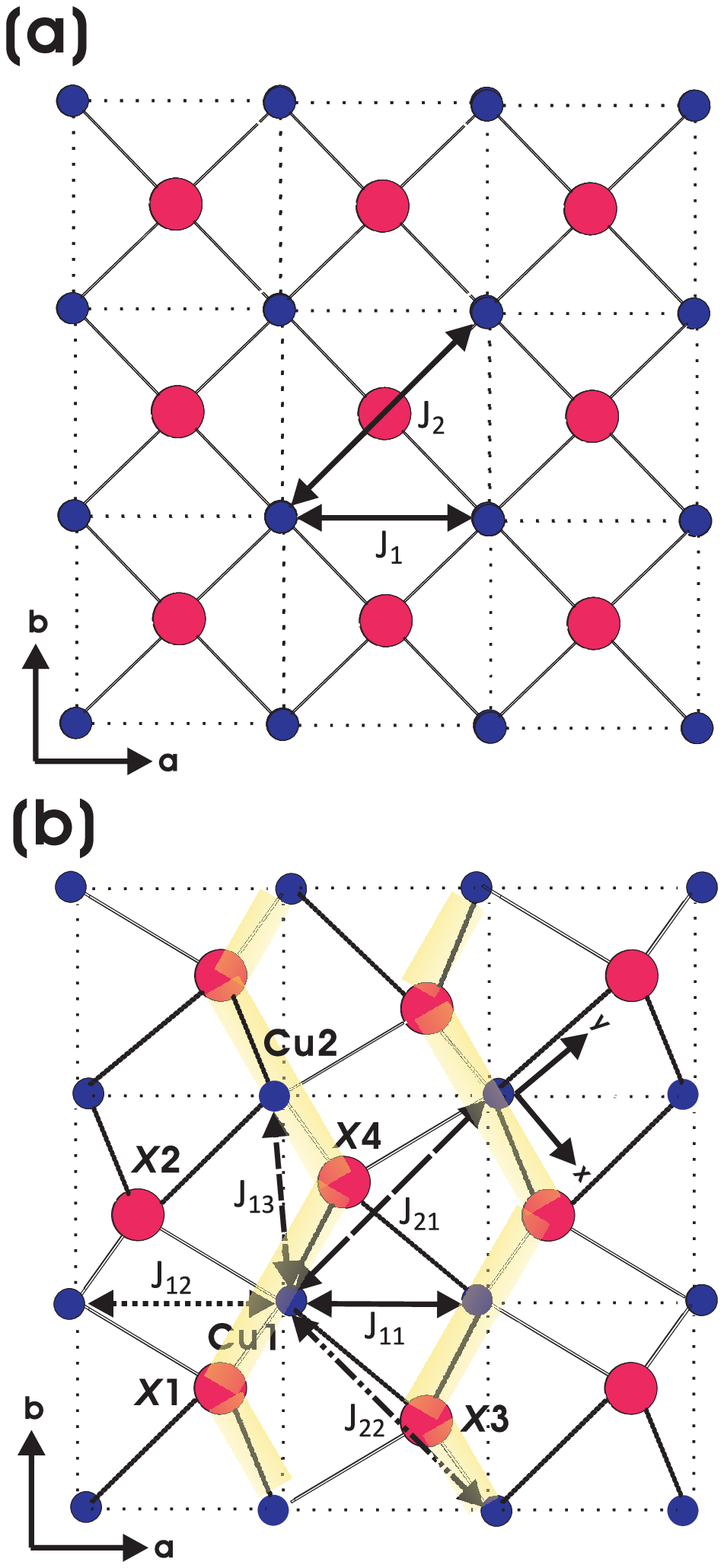}
\end{figure}

\begin{figure}[h]
        \caption{ } \label{fig3}
\includegraphics{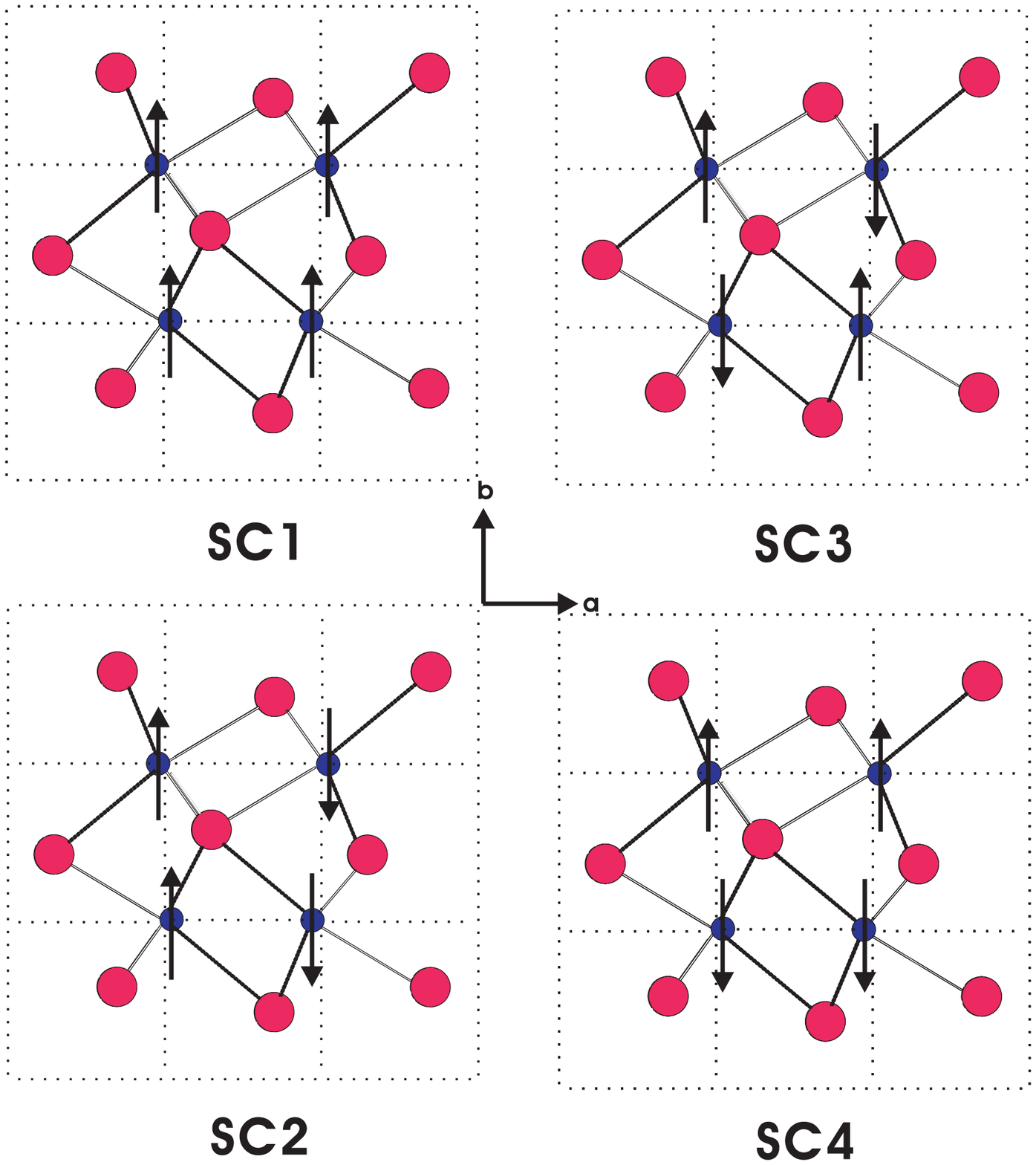}
\end{figure}

\begin{figure}[h]
        \caption{ } \label{fig4}
\includegraphics{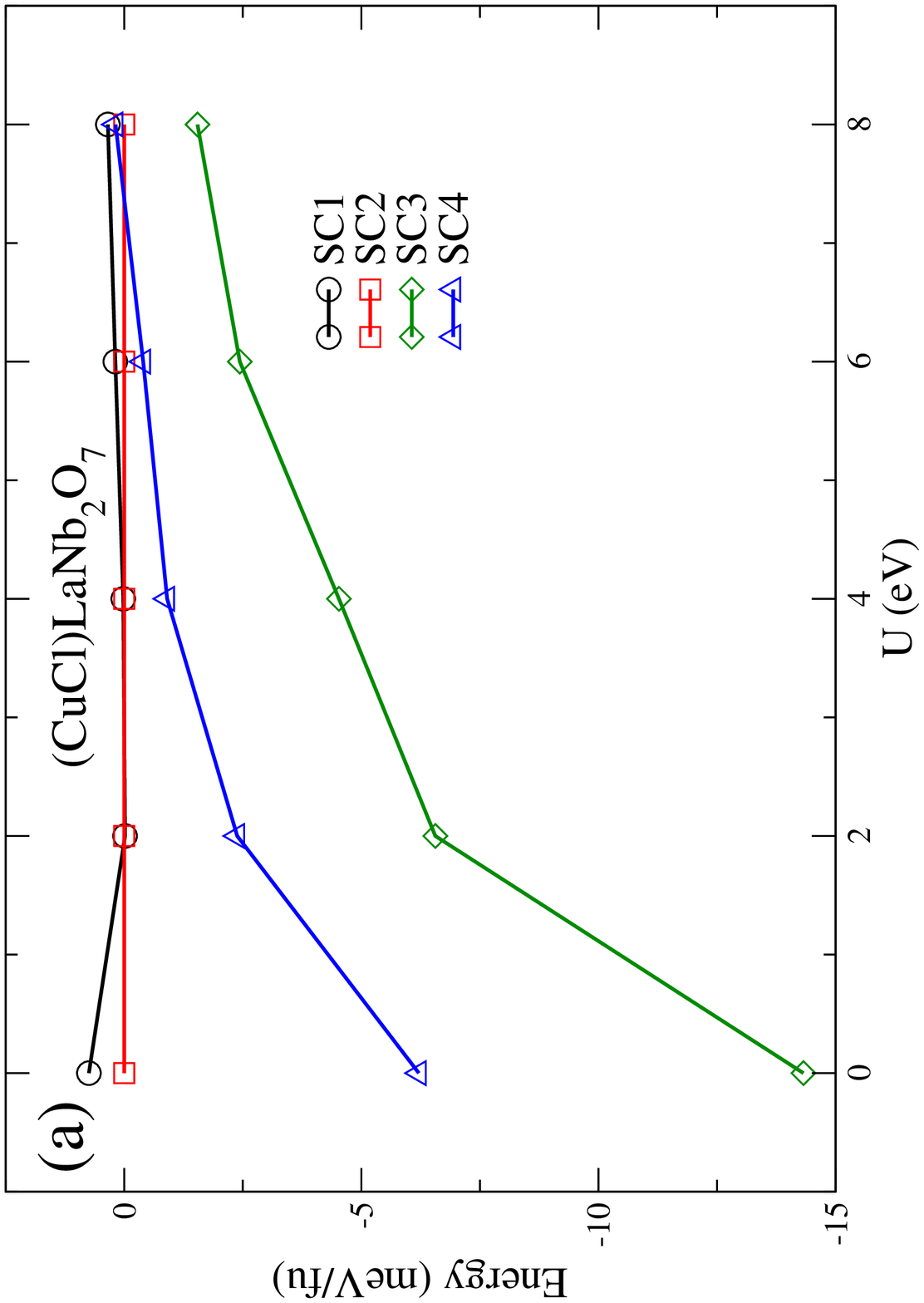}
\includegraphics{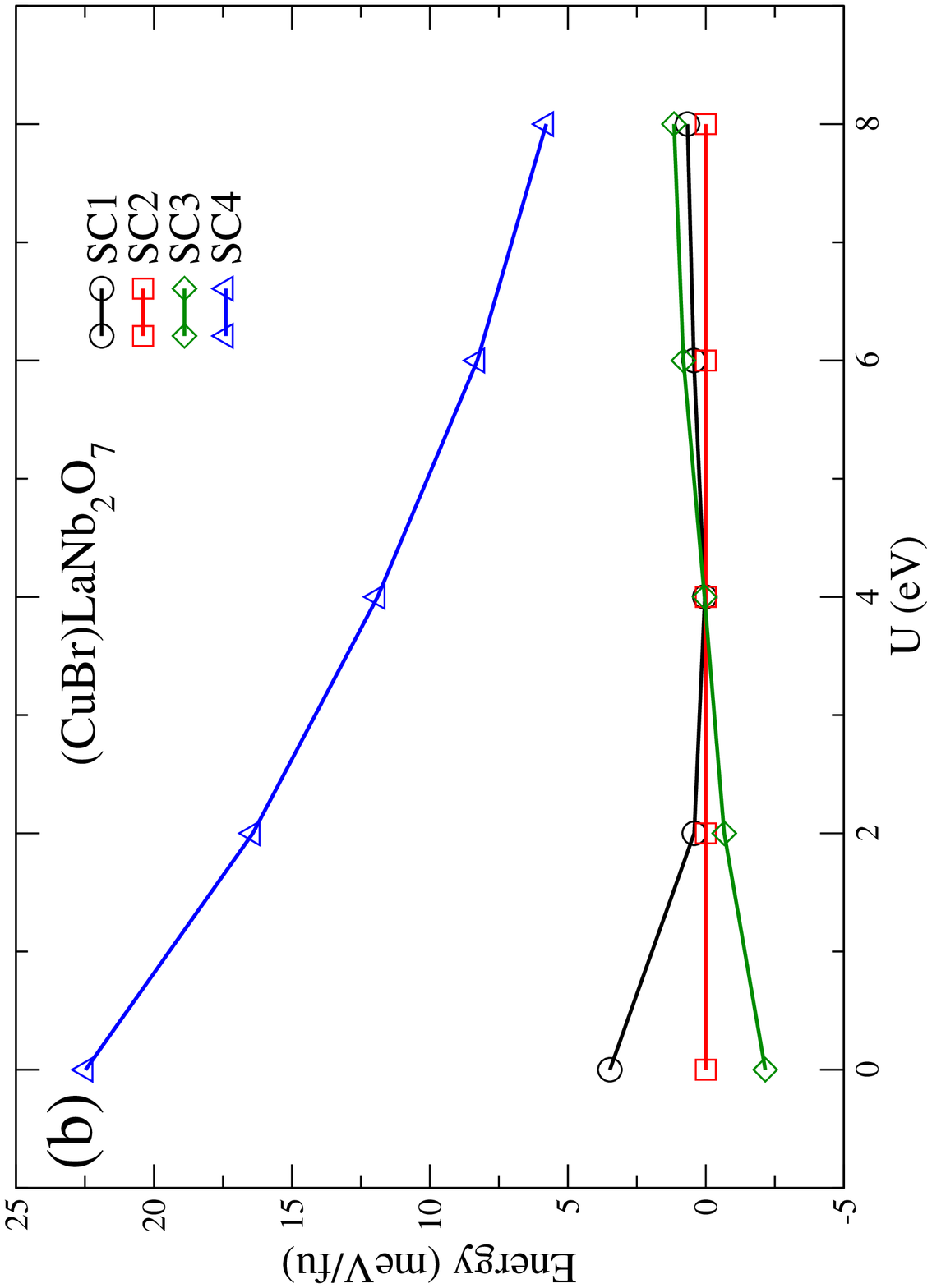}
\end{figure}

\begin{figure}[h]
        \caption{ } \label{fig5}
\includegraphics{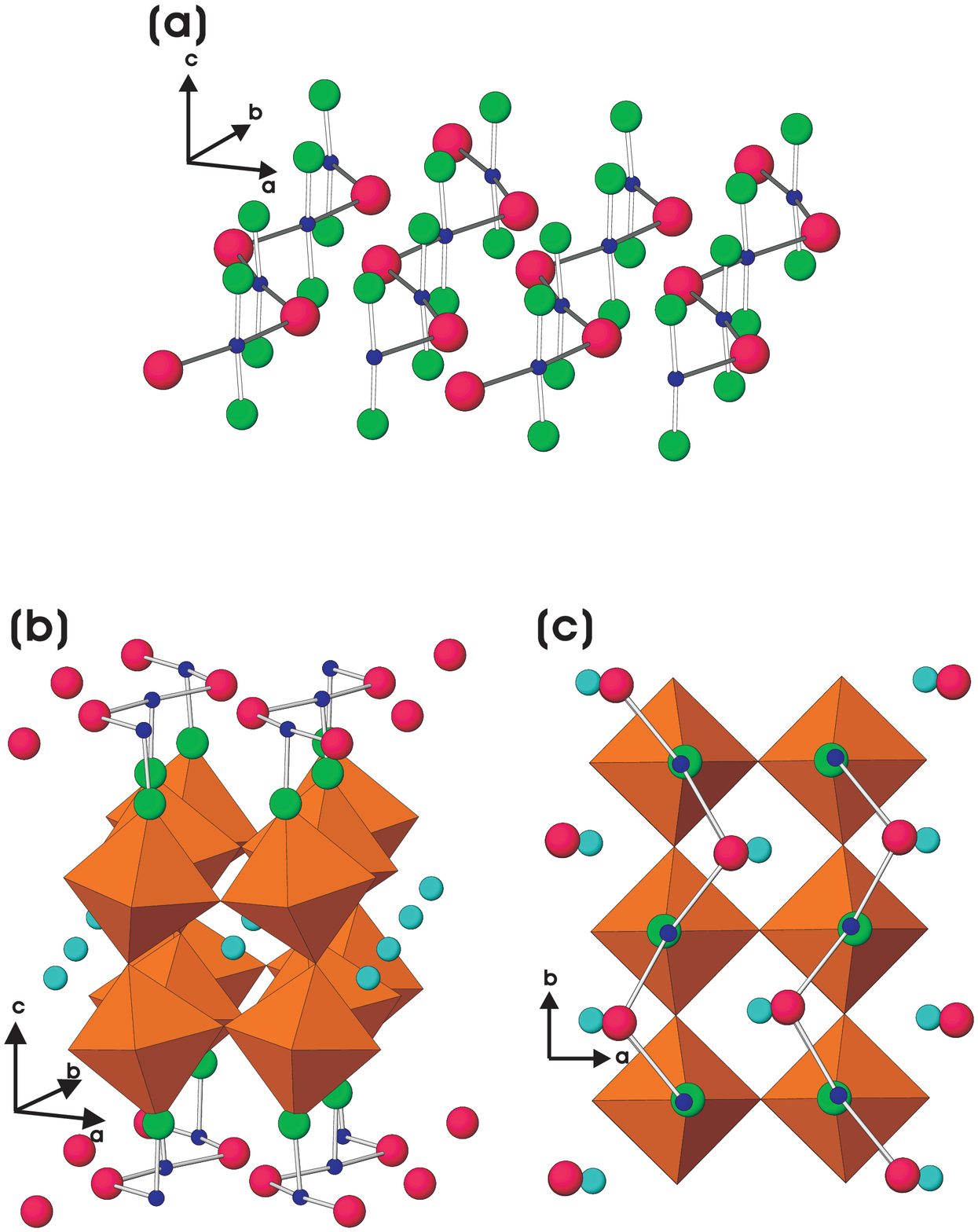}
\end{figure}

\begin{figure}[h]
        \caption{ } \label{fig6}
\includegraphics{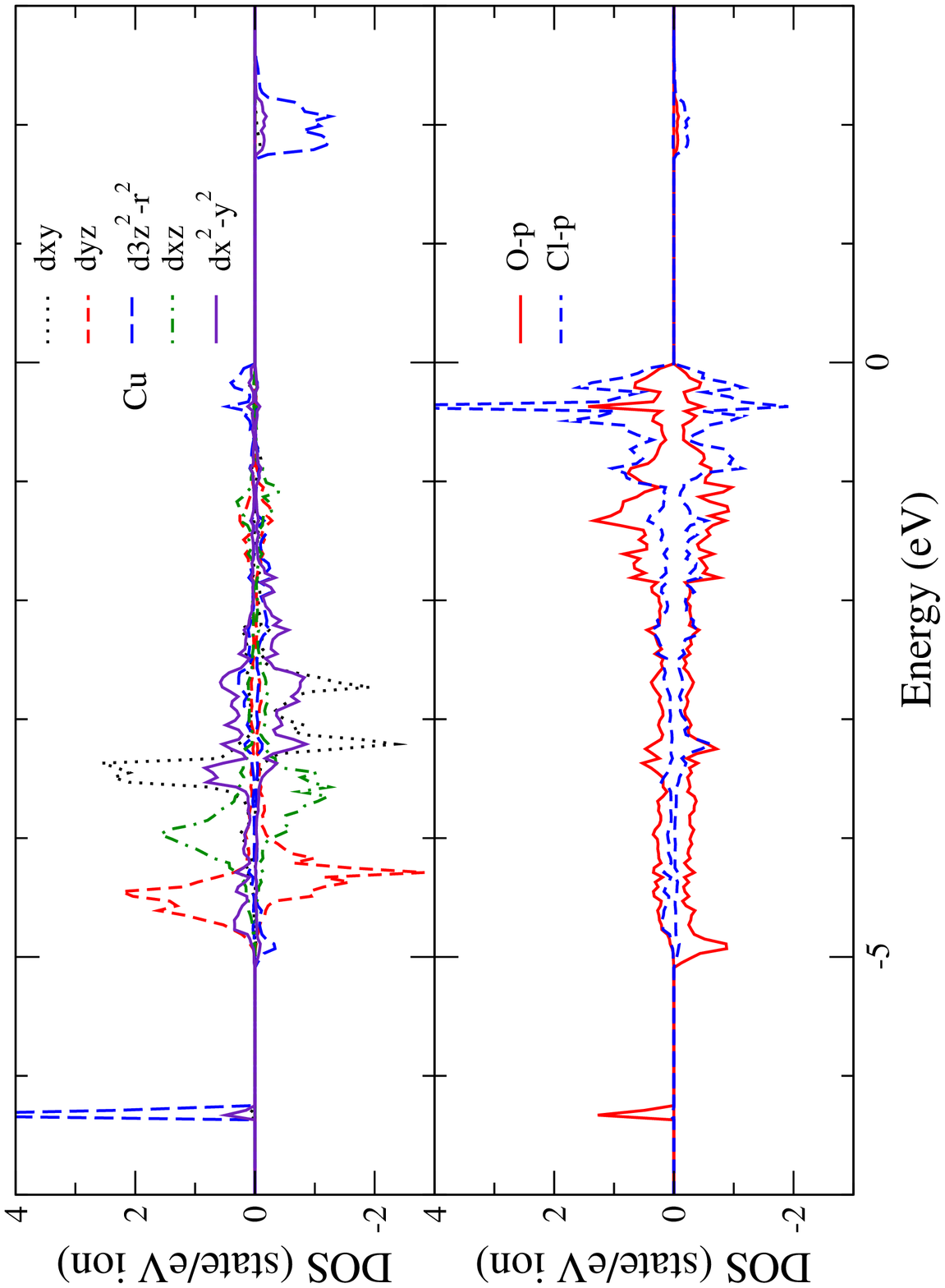}
\end{figure}

\begin{figure}[h]
        \caption{ } \label{fig7}
\includegraphics{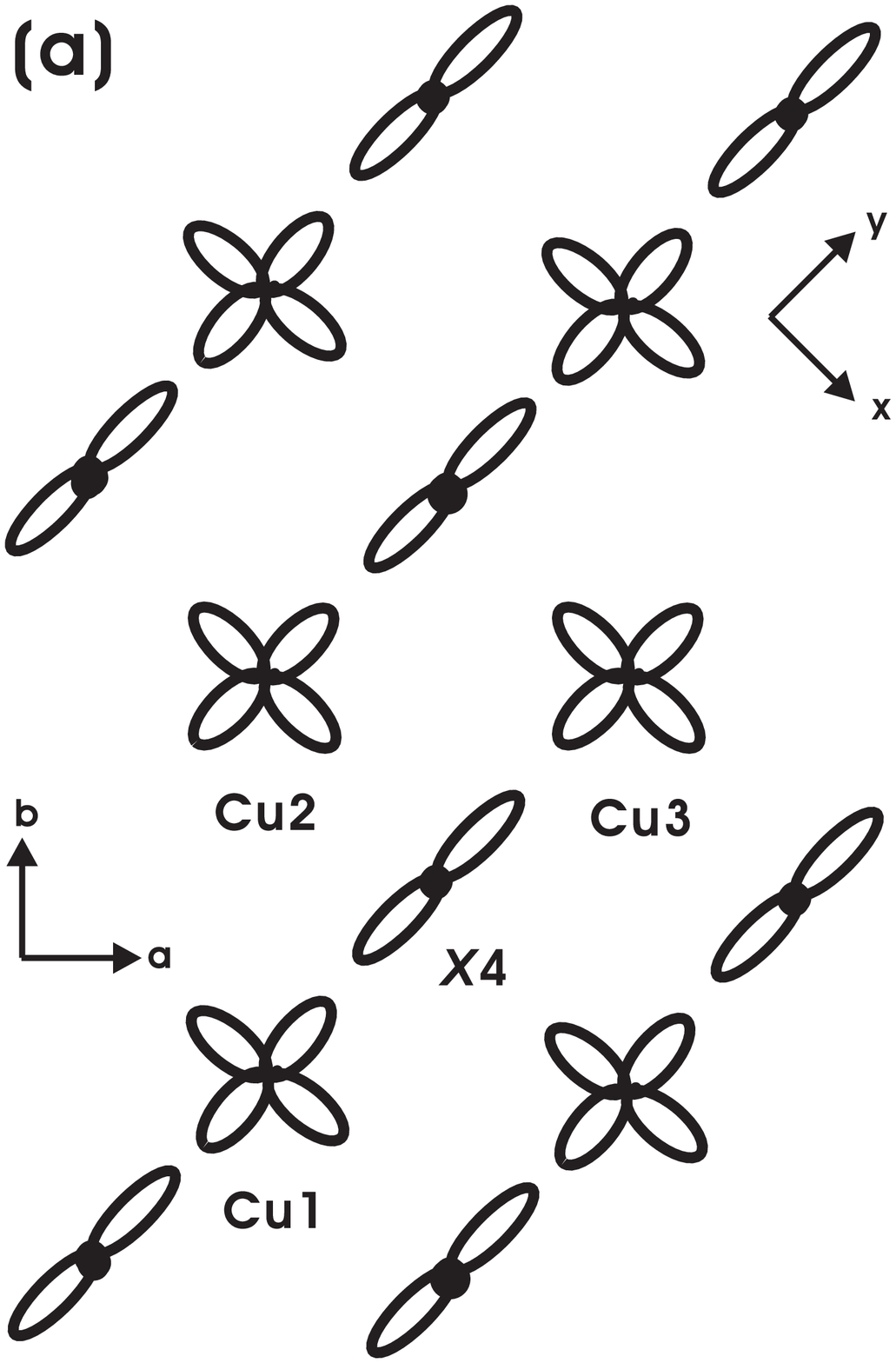}
\includegraphics{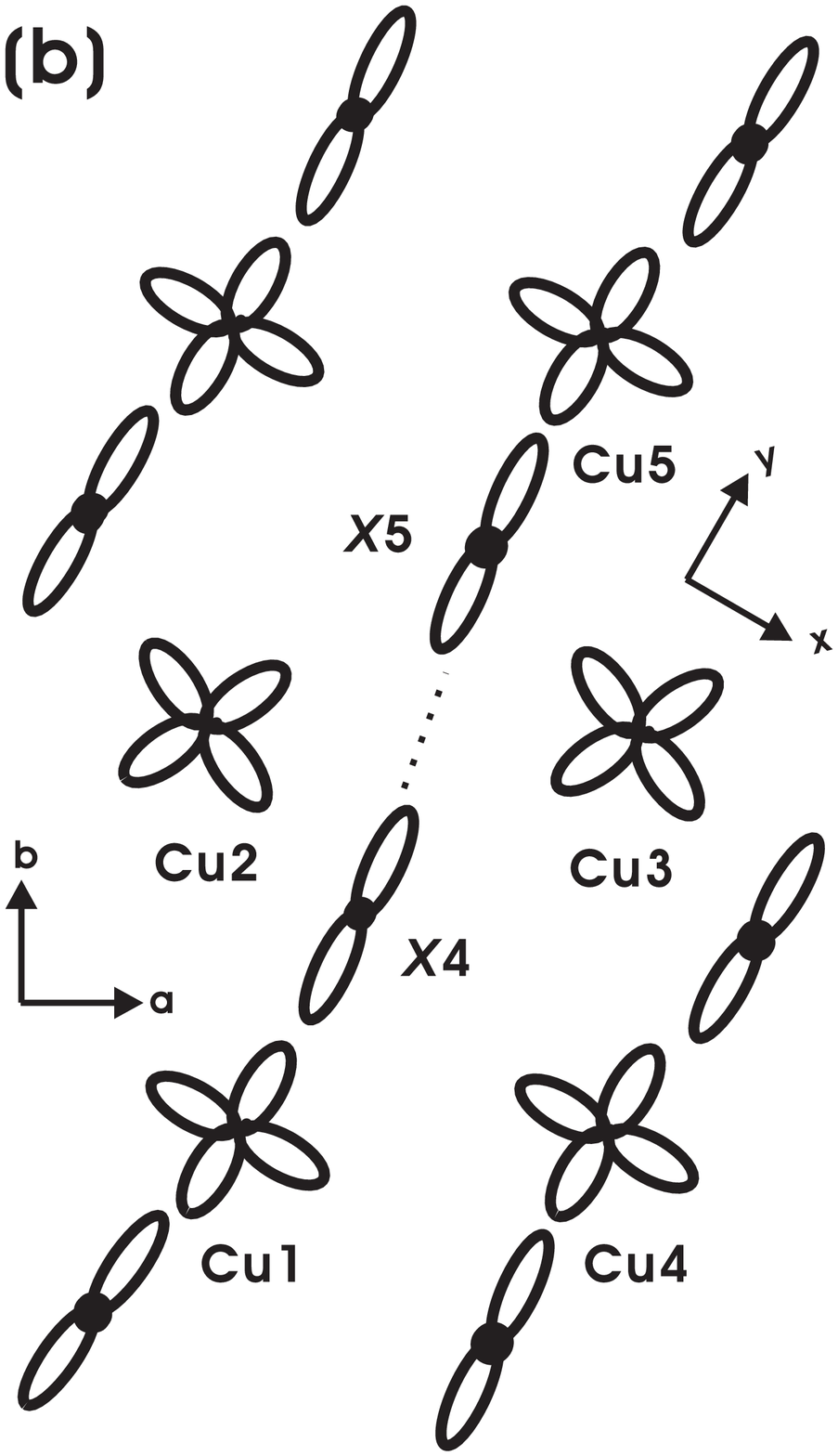}
\end{figure}

\begin{figure}[h]
        \caption{ } \label{fig8}
\includegraphics{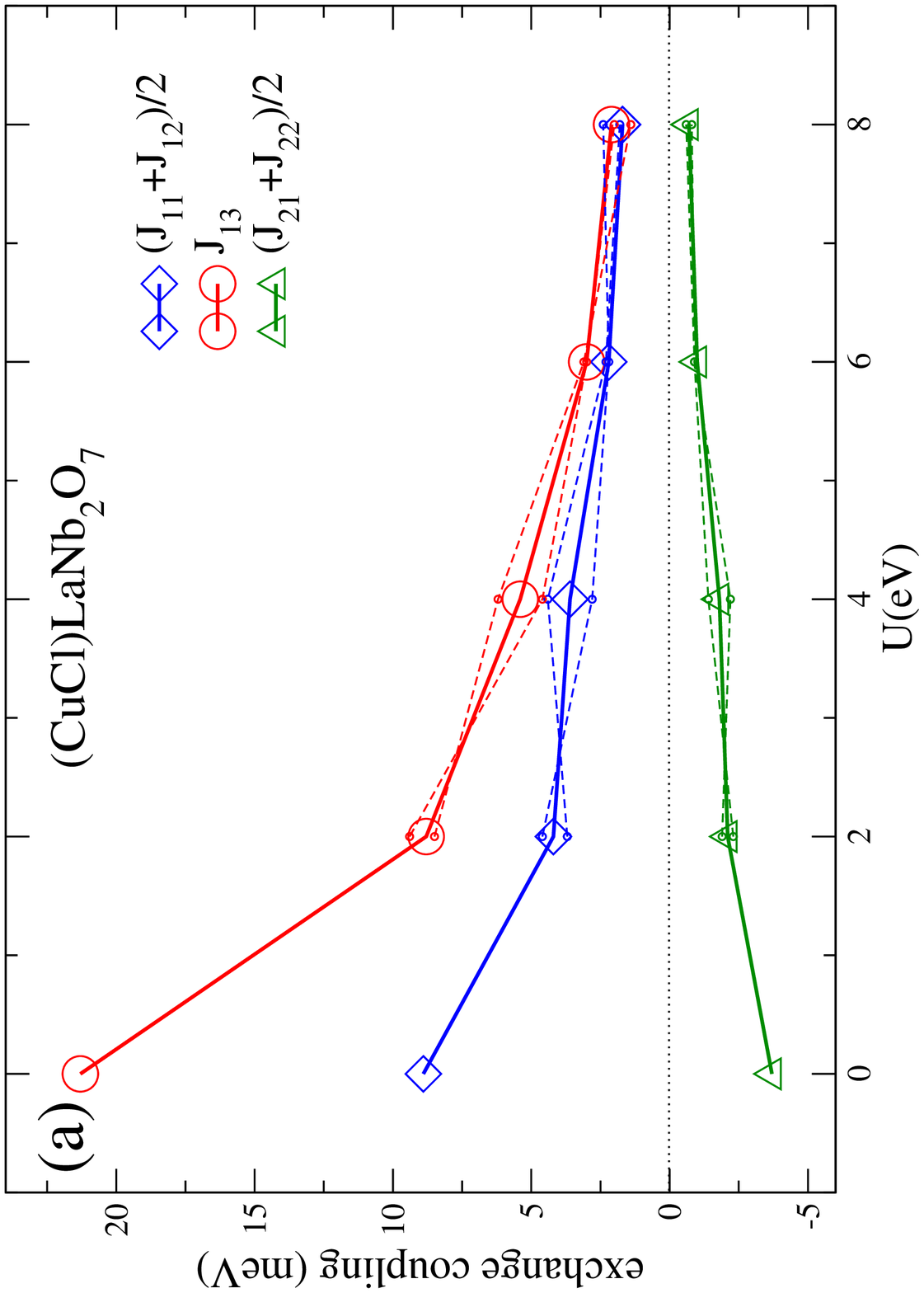}
\includegraphics{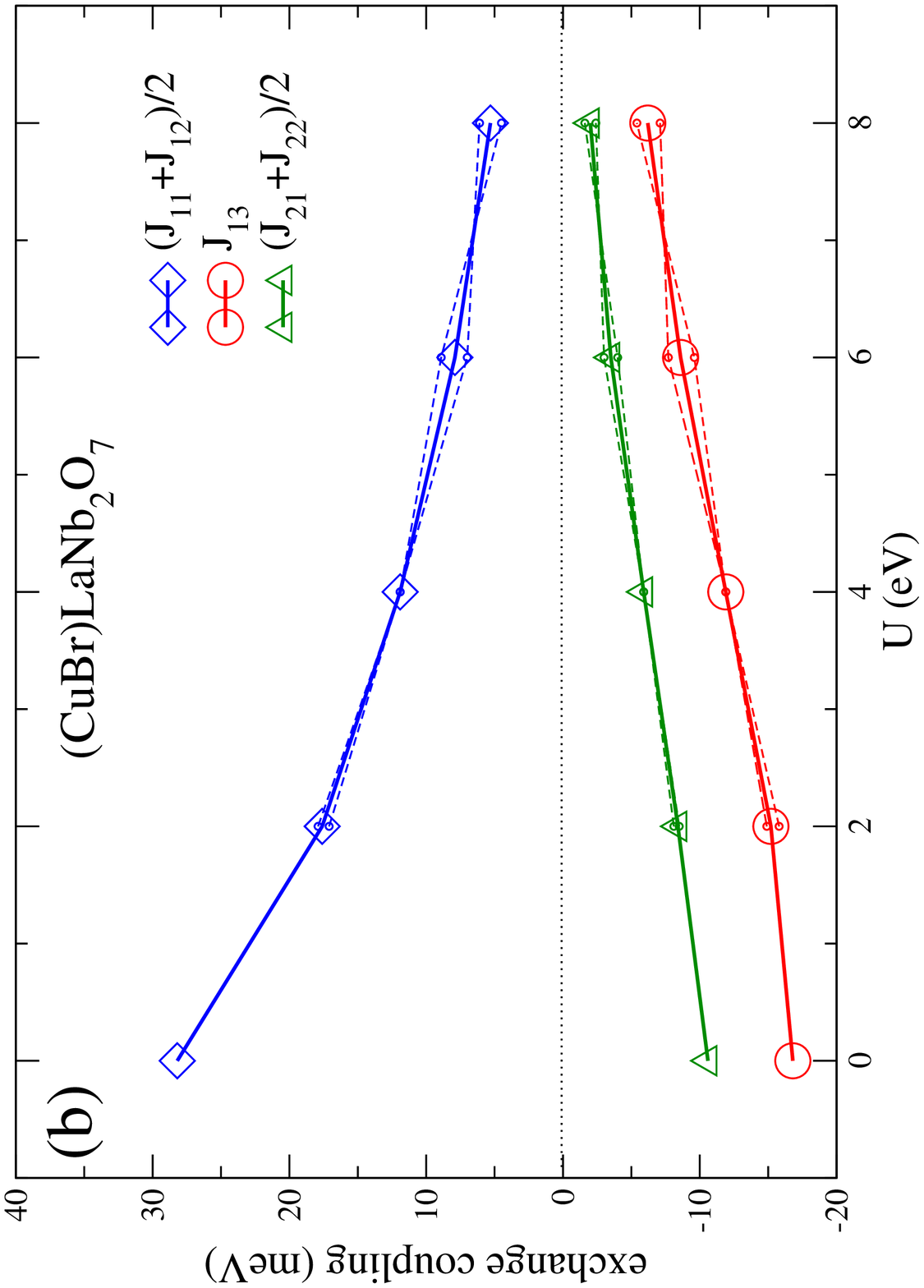}
\end{figure}

We concluded that (CuCl)LNO is still the 
AFM $J_1$ magnet with mixing FM $J_2$ interactions. 
For (CuCl)LNO, the first neighboring interactions
within the zigzag chains are AFM, and the first two neighboring couplings
between adjacent chains are AFM and FM, respectively. However, the 
replacement of Cl by Br modifies 
the first neighboring intra-chain interaction to be FM for (CuBr)LNO.
Despite the "well"-separated zigzag chains in the YY model, 
the couplings between adjacent chains
are comparable to those within the chain. The opposite 
signs of the inter- and intra-chain interactions  
in (CuBr)LNO reflect the 
spatial asymmetry and therefore the failure of the 
simple $J_1-J_2$ model for such material. All the in-plane exchange couplings 
in (CuBr)LNO are shown to be three times those in the Cl 
counterpart. It is found that the inter-plane interaction 
$J_{\perp}$ is AFM, in agreement with the experiment \cite{OKKYBNHNHKSAY}. 
The present study strongly suggests that the simple square 
$J_1-J_2$ model should
be modified to explore the magnetic property of Cu$X$LaNb$_2$O$_7$.	  
We hope the present calculations will shed light on the precise
crystallographic determination and the magnetic properties of 
Cu$X$LaNb$_2$O$_7$. 
\begin{acknowledgments}
We are grateful to P. Sindzingre for bringing these systems to our
attention. Computer resources provided by the National 
Center for High-performance Computing are gratefully acknowledged.
This work was supported by the National Science Council and 
National Center for Theoretical Sciences of Taiwan. 
\end{acknowledgments}
\newpage
%\centering
\begin{center}
\large {\bf FIGURE CAPTIONS}    \normalsize
\end{center}
Fig. 1: (Color online) Crystal structure of  
	(Cu$X$)LaNb$_2$O$_7$ ($X$=Cl, Br) in the tetragonal 
	 space group $P_4/mmm$.  \\ \\
Fig. 2: (Color online) (a) The undistorted tetragonal model and (b) the 
	distorted model proposed by Yoshida {\it et al.} \cite{YOTYIKKAY}
	for (Cu$X$)LaNb$_2$O$_7$ ($X$=Cl, Br).  
	 Large and small spheres denote $X$ and Cu ions, respectively. 
 	 The relevant exchange couplings are also indicated.  \\ \\
Fig. 3: (Color online) The four different spin configurations of Cu ions 
	considered in the present study.  
	 Large and small spheres denote Cl/Br and Cu ions, respectively.\\ \\
Fig. 4: (Color online) Total energies of (a)
	(CuCl)LaNb$_2$O$_7$ and (b) (CuBr)LaNb$_2$O$_7$ in the spin 
	configurations shown in Fig. \ref{fig3}. The energies are 
	relative to that of SC2. \\ \\
Fig. 5: (Color online)  Perspective view of (a) the  
	CuO$_2X_2$-plaquette zigzag chains and (b) tilted NbO$_6$
	octahedra of (Cu$X$)LaNb$_2$O$_7$ ($X$=Cl, Br) 
	in the space group $Pbam$. (c) Top view of (b).
	Here, the symbols for the various atomic species are the same as 
	those in Fig. \ref{fig1}. \\ \\
Fig. 6: (Color online) Orbital- and site-projected density of states 
	(DOS) of (CuCl)LaNb$_2$O$_7$, obtained by $U=6$ eV and SC2 
	 in Fig. \ref{fig3}. 
         The energy is relative to the valence-band maximum. \\ \\
Fig. 7: (Color online) Schematic plot of Cu $d(x^2-y^2)$ and $X$ 
	$p$ orbitals of (Cu$X$)LaNb$_2$O$_7$ ($X$=Cl, Br) in the 
	space group (a) $P_4/mmm$ and (b) $Pbam$.   \\ \\
Fig. 8: (Color online) The $U$-dependence of the in-plane exchange couplings of
         (a) (CuCl)LaNb$_2$O$_7$ and (b) (CuBr)LaNb$_2$O$_7$.           
 	The solid lines are calculated from SC1$-$SC4 in 
 	Fig. \ref{fig3}. The dashed lines show the uncertainty of the 
 	calculations. 
 	See the text for details. Note that the scale of exchange 
 	coupling in (a) is only half of that in (b).  
\newpage
\newpage
\newpage
\newpage
\newpage
\newpage
\newpage
\newpage
\clearpage
\begin{table}[t]
        \caption{The calculated lattice constants (\AA), relevant 
	  	 interatomic distances (\AA), and bond angles ($^{\bf o}$) of
	  	 (Cu$X$)LaNb$_2$O$_7$ ($X$=Cl, Br), 
		 obtained by $U=6$ eV and SC2 in Fig. \ref{fig3}. 
		 The latter two refer to 
	 	 those in Fig. \ref{fig2}(b). 
		 In the bottom part, the results in the undistorted 
		 tetragonal model are listed for comparison. 
		 The values in parentheses
	         are the corresponding experimental data. 
		 } \label{tab1}
\begin{ruledtabular}
                \begin{tabular}{lll}
        &(CuCl)LaNb$_2$O$_7$      &(CuBr)LaNb$_2$O$_7$        \\
\hline
$a$      &7.868    &7.889   \\
$b$      &7.883    &7.914    \\
$c$      &11.878    &11.853   \\
Cu1-$X1$      &2.39~(2.40$^{\dagger}$)    &2.54    \\
Cu1-$X2$      &3.27~(3.14$^{\dagger}$)    &3.11    \\
Cu1-$X3$      &3.29    &3.11    \\
Cu1-$X4$      &2.38    &2.52    \\
Cu1-O      &1.88~(1.84$^{\dagger}$)    &1.88    \\
$\angle$ $X$1-Cu1-$X2$      &83.6    &81.5    \\
$\angle$ $X2$-Cu1-$X4$      &87.1    &88.8    \\
$\angle$ $X4$-Cu1-$X3$      &102.7    &101.3    \\
$\angle$ $X3$-Cu1-$X1$      &86.6    &88.3    \\
$\angle$ Cu1-$X4$-Cu2      &112.2    &103.4    \\
\hline
$a^{\star}$      &3.914$~(3.884^{\dagger})$    &3.942~$(3.899^{\dagger\dagger})$  	\\
$c^{\star}$      &11.892~(11.736$^{\dagger}$)    &11.853~(11.706$^{\dagger\dagger}$) 	\\
Cu-$X$$^{\star}$  &2.77    &2.79    \\
Cu-O$^{\star}$      &1.85  &1.87    \\
                \end{tabular}
\end{ruledtabular}
$\dagger$: Ref. \cite{CKW}	\\
$\dagger\dagger$: Ref. \cite{KKZW}
\end{table}
\clearpage
\begin{table}[t]
        \caption{ The calculated atomic structural parameters of 
   		  (Cu$X$)LaNb$_2$O$_7$ ($X$=Cl and Br) in the space group 
		  {\it Pbam}, obtained by $U=6$ eV and SC2 in 
		  Fig. \ref{fig3}. $u$, $v$, and $w$ denote 
		  fractional coordinates based on the $a$, $b$, and $c$ lattice
		  constants, respectively.} \label{tab2}
\begin{ruledtabular}
                \begin{tabular}{lccccccc}
 & &\multicolumn{3}{c}{(CuCl)LaNb$_2$O$_7$}
 &\multicolumn{3}{c}{(CuBr)LaNb$_2$O$_7$} \\
\cline{3-5}   \cline{6-8}
ion &site &$u$ &$v$ &$w$ &$u$ &$v$ &$w$  	\\
\hline
Cu &$4h$  &0.2706 &0.0077 &0.5    &0.2720 &0.0057 &0.5  	\\
$X$ &$4h$ &0.4185 &0.2711 &0.5    &0.4481 &0.2713 &0.5  	\\
La &$4g$  &0.5000 &0.2622 &0      &0.5000 &0.2624 &0  		\\
Nb &$8i$  &0.2522 &0.0000 &0.1911 &0.2520 &0.9998 &0.1903  	\\
O1 &$4f$  &0.5    &0      &0.1336 &0.5    &0      &0.1334  	\\
O2 &$4e$  &0      &0      &0.1841 &0      &0      &0.1834  	\\
O3 &$8i$  &0.2498 &0.2501 &0.1522 &0.2498 &0.2499 &0.1520  	\\
O4 &$4g$  &0.2004 &0.0001 &0      &0.2004 &0.0000 &0  		\\
O5 &$8i$  &0.2830 &0.0008 &0.3417 &0.2818 &0.0000 &0.3413  	\\
                \end{tabular}
\end{ruledtabular}

\end{table}
\clearpage
\begin{table}[t]
        \caption{ The exchange couplings (meV) of 
   		  (CuCl)LaNb$_2$O$_7$ and (CuBr)LaNb$_2$O$_7$. 
		  The notation is explained in the text. 
	          $U$ (eV) is the on-site Coulomb 
		  correlation interaction.} \label{tab3}
%\scriptsize
\begin{ruledtabular}
                \begin{tabular}{lcccccccccccc}
   &\multicolumn{6}{c}{(CuCl)LaNb$_2$O$_7$}
 &\multicolumn{6}{c}{(CuBr)LaNb$_2$O$_7$} \\
\cline{2-7}   \cline{8-13}
$U$ 
&$(J_{11}+J_{12})/2$ &$J_{13}$ &$(J_{21}+J_{22})/2$ &$J_{\perp}$  
&$J_{1}$ &$J_{2}$ 
&$(J_{11}+J_{12})/2$ &$J_{13}$ &$(J_{21}+J_{22})/2$ &$J_{\perp}$  
&$J_{1}$ &$J_{2}$ 	\\
\hline
0 &8.9 &21.3 &$-3.7$ &5.1 &$-1.8$ &18.4 &28.2  &$-16.8$ &$-10.6$ &5.4 &$-2.9$ &$-1.5$  \\
4 &3.6 &5.4 &$-1.8$ &2.0 &$-3.6$ &26.9 &11.9   &$-11.9$ &$-5.9$ &2.2 &$-1.8$ &23.3  \\
6 &2.2 &3.0 &$-1.0$ &1.3 &$-2.5$ &19.4 &7.9   &$-8.6$ &$-3.5$ &1.5 &$-0.1$ &25.5  \\
8 &1.7 &2.1 &$-0.7$ &0.7 &$-1.7$ &13.5 &5.3   &$-6.2$ &$-2.0$ &0.9 &$-0.3$ &18.6  \\
                \end{tabular}
\end{ruledtabular}

\end{table}

\begin{thebibliography}{99}
\bibitem{com1}
The model in Ref. \cite{OKSAPKAY} originally proposed for (FeCl)LaNb$_2$O$_7$
(the same shown in Fig. 16(a) of 
Ref. \cite{YOTYIKKAY}) is energetically less favorable than the YY model.
\bibitem{com3}
From the viewpoint of the square $J_1-J_2$ model, these interactions should 
be regarded as the fourth-nearest-neighbor interactions.
\end{thebibliography}
\end{document}